\documentclass[english,aps,prl,preprint,showpacs]{revtex4}
\usepackage[T1]{fontenc}
\usepackage{babel}
\usepackage{graphics}

\makeatletter

\providecommand{\LyX}{L\kern-.1667em\lower.25em\hbox{Y}\kern-.125emX\@}

\makeatother
\begin{document}

\title{The tortuous behavior of lightning}

\author{Francisco Vera}

\email{fvera@ucv.cl}

\affiliation{Universidad Católica de Valparaíso, Av. Brasil 2950, Valparaíso,
Chile}

\begin{abstract}
The complex branched structure of lightning induce scientists to think
that dielectric breakdown is a very complicated phenomena, we will
show that this is not true and that simulating the structure of lightning
is an easy task, but depends strongly on boundary conditions. In this
work we will introduce a new way of understanding the origin of this
tortuous path that relies on minimizing the total energy stored in
the system.
\end{abstract}

\pacs{47.54.+r, 61.43.Hv}

\maketitle
It is well known that after charging a conductor, the electric charge
try to spread out as much as possible on its surface as a consequence
of the mutual repulsion between charges of the same sign. When this
charged conductor has a sharp end or tip, the electric field just
outside this region is bigger than the field outside other points
that are far from the tip, and when the electric field exceeds a certain
value, the dielectric medium (air for example) will break and a discharge
or spark will be initiated at the tip. A similar scenario occurs in
thunderstorms\cite{key-1}, where air currents separate negative and
positive charges producing regions of high electric fields, and when
this field exceeds the value for breaking the air a lightning stroke
is produced. The lightning discharge in thunderstorms and sparks between
charged conductors, span scales from millimeters to kilometers and
evolve following a tortuous path, forming a branched structure that
closely resembles a fractal\cite{key-2}.

\begin{figure}
{\centering \resizebox*{1\columnwidth}{!}{\includegraphics{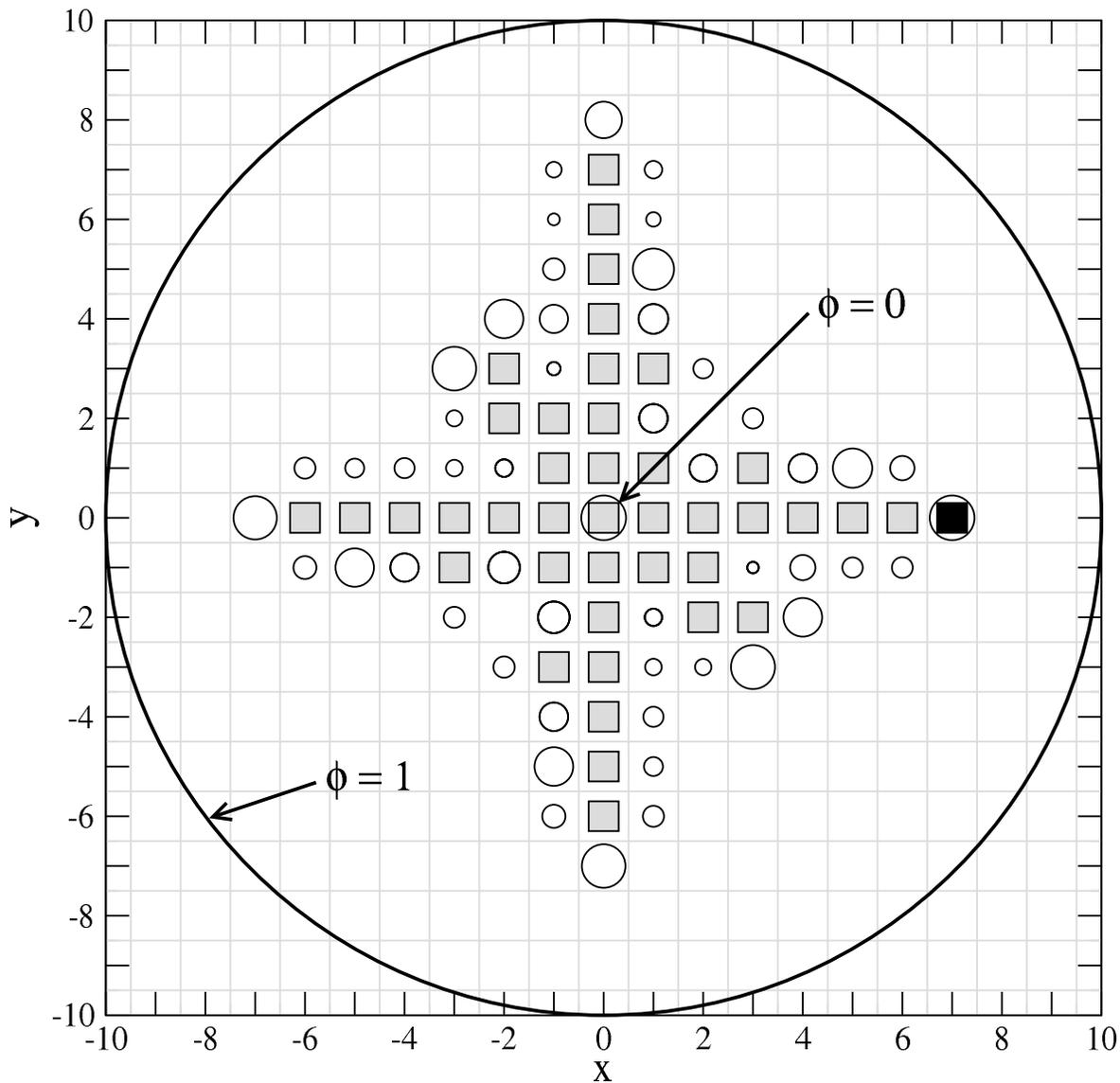}} \par}

\caption{Schematic view of the two-dimensional circular geometry used in our
calculations.}
\end{figure}

Negative cloud to ground (CG) flashes account approximately for the
40\% of all discharges in natural lightning, there are also: intra-cloud,
cloud-to-cloud and cloud-to-air discharges\cite{key-3,key-4}. A typical
negative CG flash last for about half a second and can lower an electrical
charge of some tens of coulombs. This flash can include three or four
current pulses called strokes that last about a millisecond and are
separated in time for several tens of milliseconds. These CG lightning
discharges are initiated by a downward-moving negative stepped leader,
each step having a typical duration of 1 \( \mu  \)s, tens of meters
in length and pauses between steps of 20 to 50 \( \mu  \)s. In its
way from cloud to ground the stepped leader produces a typical downward-branched
structure, the average velocity of propagation is about 2 \( \times  \)10\( ^{5} \)m/s,
the average leader current is between 100 and 1000 A, and the potential
difference between the lower portion of the leader and the Earth has
a magnitude in excess of 10\( ^{7} \)V. As the tip of this negative
leader nears ground, the electric field at sharp objects on the ground
increases until it exceeds the breakdown strength of air. At that
time, one or more upward-moving discharges are initiated from those
points, and an attachment with the downward-moving leader occurs some
tens of meters above ground. Then the first return-stroke heats the
leader channel to a peak temperature near 30,000 K producing thunder,
and after several strokes flashing in this channel disappear.

When one is faced with the problem of simulating lightning the first
thing to do is to simplify the geometry of the problem. The most used
geometries are: a circular conductor (outer electrode) plus a central
electrode (see Fig. 1), and an elongated rectangular region with a
small electrode in one end and the other end acting as the other electrode.
Then an electric potential is applied to each electrode, the difference
of potential must be big enough to reach the necessary electric field
to break the dielectric medium. All conclusions obtained in these
two-dimensional geometries can be readily applied to the real 3-dimensional
case.

In the theoretical study of the evolution of the discharge channel,
the simplest possibility is to expand the inner electrode towards
the regions of maximum electric fields. For example, the evolution
in the circular geometry proceed in the following way: first, implement
boundary conditions for the potential \( \phi  \) in the central
and outer electrodes; second find the region of higher electric field
(this will be a neighbor of the inner electrode); third, extend the
inner electrode as to include the region of high electric field; fourth,
calculate again the electric field using the new boundary conditions
(the inner electrode has a different form but the potential is maintained
constant) and repeat this procedure until the outer electrode is reached.
This necessarily gives as a result a straight line between both electrodes
as a consequence of the tip effect mentioned previously.

There exists several models in the literature that attempt to find
the tortuous structure of lightning, based in procedures similar to
the one mentioned in the previous paragraph. The presently accepted
model of lightning\cite{key-5,key-6,key-7} was developed by Niemeyer,
Pietronero and Wiesmann in 1984, and follow these steps but includes
a stochastic term that weights a probability that is a function of
the value of the local electric field, this model is known as the
Dielectric Breakdown Model (DBM) and produces a branched structure
whose fractal dimension is similar to the ones obtained experimentally
for the same geometry.

In this work we will show that it is possible to obtain a much more
realistic branched structure of lightning that follow from a deterministic
treatment, that only relies in minimizing the total energy stored
in the system and changing locally the dielectric permitivity of the
medium (not the geometry of the inner electrode) at each step of iteration.

Before proceeding to explain our model of lightning, we must review
some physical concepts that will help us to understand the underlying
simple mechanisms acting on the system to produce the branched structure.

The parallel plate capacitor is a very simple electrical device that
has all the necessary ingredients to understand the physical concepts\cite{key-8}
under work in our problem at hand. This device consists of two parallel
conductive plates (of area A) separated by a short distance (d) and
filled with a dielectric material of dielectric permitivity \( \varepsilon  \).
The energy stored in the electric field can be obtained from\begin{equation}
\label{eq1}
U=\frac{1}{2}\int \varepsilon (\mathbf{E}\cdot \mathbf{E})dV.
\end{equation}
 If we disconnect the battery after charging the capacitor, the charge
in this device will be held constant, and the energy stored in the
electric field (the total energy) will be proportional to d/\( \varepsilon  \).
Physical systems always evolve trying to reduce the total stored energy\cite{key-9},
this implies a force between the plates trying to reduce the distance
d, and if you insert a slab of a material having a dielectric permitivity
\( \varepsilon  \)' greater than \( \varepsilon  \), the slab will
be pulled into the capacitor. Now we came to a key point for our model,
if instead of maintaining Q constant the potential difference is maintained
constant, the energy stored in the electrical field is proportional
to \( \varepsilon  \)/d. In experiments at constant Q or V there
are forces trying to reduce d and to increase \( \varepsilon  \).
From this we learn that in experiments at constant V, U is not the
total energy of the system because the system evolves trying to increase
U. There is a missing energy term that comes from a rearrangement
of charges in the wires to maintain V constant. When this term is
introduced, the total energy is again proportional to d/\( \varepsilon  \).
In short, we can study systems at constant V by letting them to evolve
towards regions of higher U (or C).

The electrostatic energy U can be obtained as follows: First, impose
the boundary conditions: \( \phi  \)=0 in the inner electrode and
\( \phi  \)=1 in the outer electrode. Second, impose a fixed constant
value for the dielectric permitivity \( \varepsilon  \) in all the
region between the electrodes. Third, solve the poisson equation\begin{equation}
\label{eq2}
\nabla \cdot (\varepsilon \nabla \phi )=-\rho ,
\end{equation}
obtaining the potential \( \phi  \) in the region between the electrodes,
in our case we do not consider the presence of free charges in this
region and we set the charge density \( \rho  \) to zero. Fourth,
obtain the electric field by taking the gradient of this potential,
and using Eq. 1, obtain the electrostatic energy.

In the case of cylindrical symmetry, the electric field and the energy
U can be obtained analytically. After the first step in the evolution
of the discharge channel the system loose this symmetry, and we have
to rely to numerical methods for obtaining the energy U.

Now we will study the dielectric breakdown in the circular geometry
of Fig. 1. We consider a two-dimensional square lattice, where the
central point is the inner electrode and the outer electrode is modeled
as a circle. The boundary conditions, \( \phi  \)=0 in the inner
electrode and \( \phi  \)=1 in the outer electrode, are maintained
trough all the steps in our simulation. The first step is to assign
a fixed value \( \varepsilon  \) to the dielectric permitivity of
each lattice point between the electrodes and set the discharge channel
to the central point. The second step is to obtain the energies U
of the system after changing the dielectric permitivity in one of
each neighbor of the discharge channel to a greater value \( \varepsilon  \)',
these energies are compared and the neighbor providing the bigger
energy value is added to the channel. This last step is repeated until
the channel reach the outer electrode.

There are several points worth to mention about our model of lightning:
To maintain the model simple, channel evolution through the diagonals
are not permitted. Tip and screening effects are obviously incorporated,
but they are not the only essential points in the model. Changing
the local dielectric permitivity in not as drastic as changing the
geometry of the inner electrode. The local change in permitivity is
a key point of our model, providing the mechanism for the system to
develop a branched structure in the evolution towards a configuration
that minimizes the total stored energy. Our simulated discharge channels
for small lattices don't look very realistic, but obtaining better
results using bigger lattices, would require to implement sophisticated
numerical methods.

Fig. 1 shows a discharge channel for a 20x20 square lattice after
40 iterations. The inner electrode is represented by the central site
and the outer electrode is represented by a discretized version of
the big circle. For each different configuration, the numerical solution
of Eq. 2 was accepted when the numerical residual was less than 10\( ^{-4} \),
the values for the dielectric permitivity outside the channel was
\( \varepsilon  \)=1 and inside the channel was \( \varepsilon  \)'=5.
The filled boxes represent the discharge channel (sites where the
permitivity is \( \varepsilon  \)'). The circles show all possible
sites where the channel can evolve, the diameter of each of these
circles represent the value for the energy U of the system in case
this site is added to the channel. The circle at the extreme right
of the figure, having the black box inside, represent the site giving
the biggest contribution to the system energy and is the next step
in the evolution of this discharge channel.

\begin{figure}
{\centering \resizebox*{1\columnwidth}{!}{\includegraphics{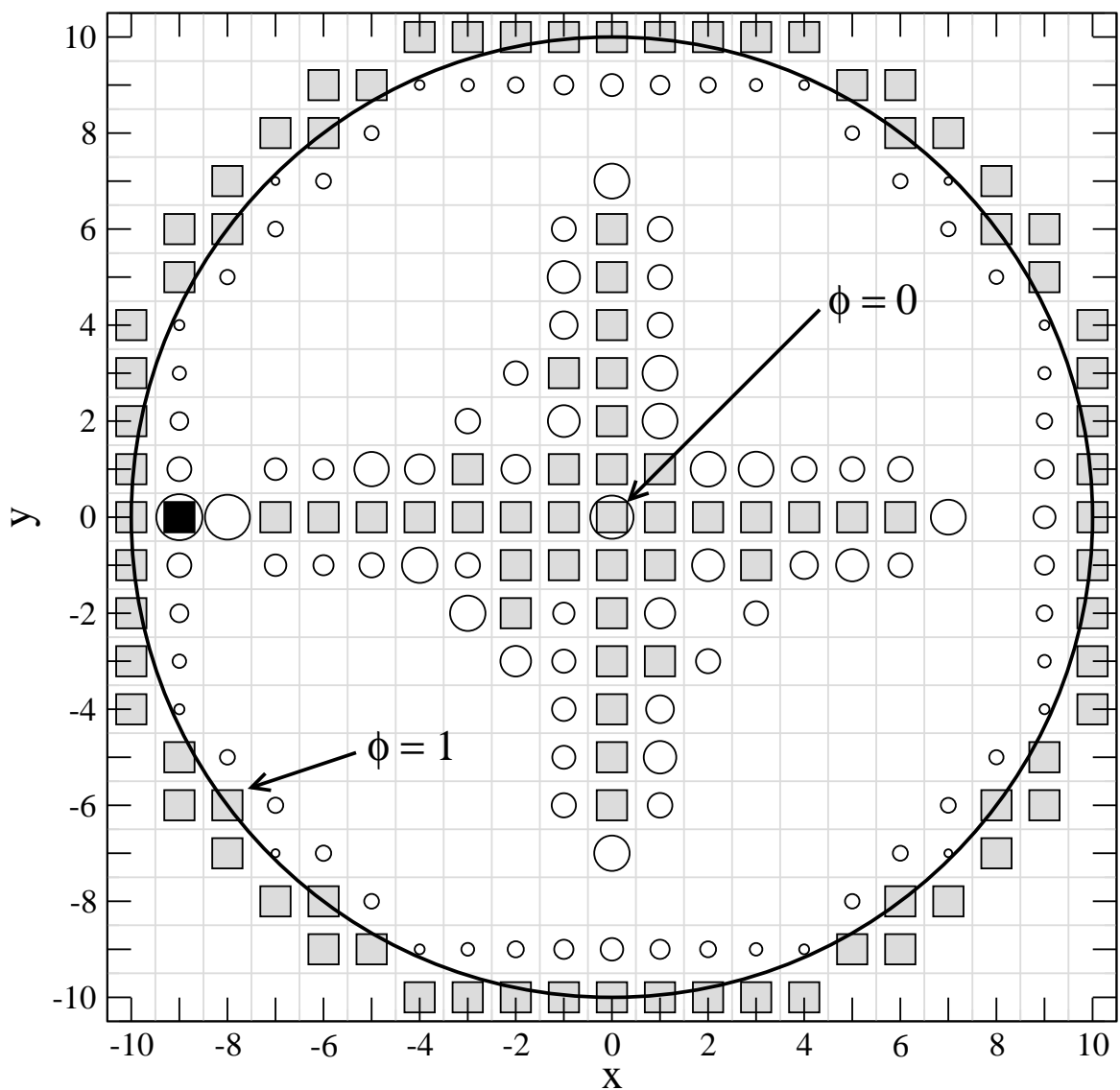}} \par}

\caption{Discharge channel, for a 20 x 20 square lattice, showing the attachment
between the return branch (outer gray boxes) and the main branch (inner
gray boxes).}
\end{figure}

The evolution of the discharge channel shown in Fig. 1 shows several
important aspects that are not present in other models of lightning:
The central site grows at the beginning evolving towards each neighbor,
forming a central cross. Then each of the four tips is prolonged forming
a bigger central cross (only one of these tips would grow if only
the tip effect were at work). This central cross increase size until
some branches develop. Although diagonals are not permitted, the channel
performs a zigzag evolution to form branches from the central site
along the diagonals.

Upward-moving discharges initiated from earth attach with the downward-moving
leader in real lightning. We extended our model to consider two initial
branches: the main branch coming from the central lattice site and
the return branch coming from the outer electrode. Apart from considering
the possible evolution of the system by extending the return branch,
no other changes were made to the numerical implementation of our
model. In Fig. 2 we show the evolution for the same system as the
one used for obtaining Fig. 1, but including the possibility of a
return branch and accepting the numerical solution of Eq. 2 when the
numerical residual was less than 10\( ^{-2} \). The return branch
begin to evolve, after 36 steps of evolution of the main branch, by
extending the return branch towards the site having the black box
inside (at the extreme left of the figure). The site between this
point and the main branch is the next step in the evolution, completing
the path for this discharge channel from the inner electrode towards
the outer electrode. Although the configuration of the inner part
of the discharge channel is the same for Fig. 1 and Fig. 2 (rotated),
its evolution after the attachment is different. Obtaining the return
branch in our numerical simulations is a really satisfactory finding,
as nobody dreamed in the possibility of obtaining this attachment
using any of the present models of lightning.

Presently, there is a great scientific interest in the realistic modeling
of lightning\cite{key-10} for: improving models of storms for climate
studies; helping in the prevention of aircraft, spacecraft, and other
accidents; etc. The numerical implementation of our model is similar
(although conceptually different) to others models of lightning, making
easy to others groups to check and extend our results. It would be
interesting to test if the lightning channel is already formed before
any electrical current is transported by the channel, this could be
achieved in experiments using optical methods for measuring the differences
of refraction index in the system.

The physical mechanism underlying our work is not limited to explain
the branched structure of lightning, and we expect that our work will
help to understand the origin of branched structures in many other
physical systems.


\begin{thebibliography}{10}
\bibitem{key-1}B. Franklin, {}``Experiments and Observations on Electricity Made
at Philadelphia'' \emph{}(E. Cave, London, 1774).
\bibitem{key-2}B. Mandelbrot, {}``Fractals: Form, Chance and Dimension'' (Freeman,
San Francisco, 1977).
\bibitem{key-3}M. A. Uman, {}``The Lightning Discharge'' (Academic Press, San Diego,
CA, 1987).
\bibitem{key-4}M. A. Uman and E. P. Krider, Science \textbf{246}, 457 (1989).
\bibitem{key-5}L. Niemeyer, L. Pietronero, and H. J. Wiesmann, Phys. Rev. Lett. \textbf{52},
1033 (1984).
\bibitem{key-6}A. Erzan, L. Pietronero, and A. Vespignani, Rev. Mod. Phys. \textbf{67},
545 (1995).
\bibitem{key-7}R. Cafiero, A. Gabrielli, M. Marsili, L. Pietronero, and L. Torosantucci,
Phys. Rev. Lett. \textbf{79}, 1503 (1997).
\bibitem{key-8}J. C. Maxwell, {}``A Treatise on Electricity and Magnetism'' (Oxford
University Press, 1873).
\bibitem{key-9}R. P. Feynman, R. B. Leighton, and M. Sands, {}``The Feynman Lectures
on Physics'', (Addison Wesley, 1964) vol. 2.
\bibitem{key-10}E. R. Mansell, D. R. MacGorman, C. L. Ziegler, and J. M. Straka, J.
Geophys. Res. \textbf{107}, 10.1029/2000JD000244 (2002).\end{thebibliography}
\end{document}